\documentclass[showpacs,aps,amssymb]{revtex4}
\usepackage{graphicx}
\usepackage{psfig}
\usepackage{graphics}
\usepackage{amsmath}

\begin{document}
\title{Bound states in the dynamics of a dipole in the presence of  a conical defect}

\author{C. A. de Lima Ribeiro}
\affiliation{Departamento de F\'{\i}sica - Universidade Estadual de Feira de Santana\\
BR 116-Norte, Km 3, 44.031-460 - Feira de Santana, BA, Brazil}

\author{Claudio  Furtado}
\affiliation{Departamento de
 F\'{\i}sica-Universidade Federal da Para\'{\i}ba, Caixa Postal 5008, 58.051-970 - Jo\~ao Pessoa, PB, Brazil}
 
\author{Fernando Moraes}
\affiliation{Laborat\'orio de F\'{\i}sica Te\'orica e Computacional\\Departamento de
 F\'{\i}sica-Universidade Federal de Pernambuco\\50.670-901 - Recife, PE, Brazil}
\affiliation{Departamento de
 F\'{\i}sica-Universidade Federal da Para\'{\i}ba, Caixa Postal 5008, 58.051-970 - Jo\~ao Pessoa, PB, Brazil}

\begin{abstract}
In this work we investigate the quantum dynamics  of an electric dipole   in  a $(2+1)$-dimensional conical spacetime.   For specific conditions, the Schr\"odinger equation is solved  and bound states are found  with the energy spectrum and eigenfunctions  determined. We find that the bound states spectrum extends from minus infinity to zero with a point of accumulation at zero. This unphysical result is fixed when a finite radius for the defect is introduced.
\end{abstract}


\maketitle

Lower-dimensional gravity, besides its pedagogical role \cite{1}, has important applications in condensed matter physics: elastic solids with defects can be mapped into three-dimensional gravity with torsion \cite{katanaev}. Within this framework, we have studied \cite{3,jpa:har,moraes} a few cases concerning classical and quantum effects in solid media containing topological defects such as disclinations and dislocations (a disclination in a solid is analogous to a cosmic string in spacetime). In particular, the scattering states of a single electric dipole in a (2+1)-dimensional conic space-time was investigated in \cite{ribeiro}. From a different point of view, the dipole dynamics was also studied in \cite{spavieri}. An important issue in this class of problems is the question  of the self-interaction of a point charge in the background space of a topological defect \cite{linet,smith,6d,7d}. Due to its simplicity the  cosmic string  provides a prototype spacetime for the investigation of the non-local influence of gravity on matter fields. The effect is usually studied in the spacetime of a single cosmic string, but multiple cosmic strings and multiconical space-time have also been considered \cite{valdir}.  For simplicity, we restrain our study to the (2+1)-dimensional section of the cosmic string metric
\begin{equation}
ds^{2}=-c^2dt^2+d\rho^{2}+\frac{\rho^{2}}{p^{2}}d\theta^{2},
\end{equation}
in polar coordinates. The linear mass  density $\mu$ is related to the parameter $p$ by  $p=(1-4G\mu)^{-1}>1$, where $G$ is the gravitational constant. This dimensional reduction can be justified by invoking translational invariance along the string axis.

In this letter, the study of the bound states of an electric dipole to a $(2+1)$-dimensional conical singularity is analyzed. Treating the self-interaction by a classical field theory approach \cite{grats},  the self-energy is obtained and so is the Schr\"odinger equation in this background. By solving the Schr\"odinger equation the energy spectrum is found. 

The effective interaction between the dipole and the defect is the electromagnetic self-energy of the dipole due to the non-flat background created by the defect. For instance, a point charge in a static gravitational field experiences an electrostatic force due to the deformation of its electric field lines produced by the particular geometry of space-time \cite{witt}. The net effect is a self-force on the charge. A straightforward method to  calculate the self-energy in a $(2+1)$-dimensional space-time has been developed by Grats and Garcia \cite{grats}. The regularized self-energy is therefore found to be
\begin{equation}\label{07}
U=\frac{(p^{2}-1)}{48\pi}\frac{D^{2}\cos(2\phi)}{\rho^{2}},
\end{equation}
where $\phi$ is the angle between a vector directed from the singularity and the dipole direction and $D$ is the dipole moment. 

Notice that a minimum for the self-energy is obtained when  $\phi=\frac{\pi}{2}$. From this it is seen that deflections from equilibrium position leads to a  restorative force which tends to align the dipole perpendicular  to the vector directed from the singularity. This rotational degree of freedom will have a small contribution for the overall dynamics of the dipole, in particular if we are looking at low or null rotational angular momentum states. Therefore, from now on we will not consider this degree of freedom; that is, we take 
\begin{equation}
U=-\frac{(p^{2}-1)D^{2}}{48\pi \rho^{2}}.
\end{equation}
We thus have an attractive inverse square potential. This kind of potential  has attracted substantial attention recently. It was  extensively studied by Case \cite{case} in 1950 and more recently by a number of authors \cite{canal,gupta}. Recently, Audretsch \textit{et al.} \cite{audretsch} studied a similar potential in the context of the  Aharonov-Bohm scattering problem where the bound states are possible solutions.

With the kinetic energy  given in terms of the Laplace-Beltrami operator 
\begin{eqnarray}
\nabla^{2}=\frac{1}{\rho}\partial_{\rho}(\rho\partial_{\rho})+\frac{p^{2}}{\rho^{2}}\partial_{\theta\theta},
\end{eqnarray}
we arrive at the following Schr\"odinger equation
\begin{eqnarray}
\left[\frac{1}{\rho}\partial_{\rho}(\rho\partial_{\rho})+\frac{p^{2}}{\rho^{2}}\partial_{\theta\theta}+\frac{2m\mathcal{E}}{\hbar^{2}}+\frac{(p^{2}-1)d^{2}}{48\pi \rho^{2}}\right]\Psi(\rho,\theta)=0,
\end{eqnarray}
where $d$ is the normalized dipole moment, given by $d^2=\frac{2m}{\hbar^2}D^2$. 

Considering  rotational symmetry we write the solution in the form  
\begin{equation}
\Psi(\rho,\theta) = R(\rho)\exp{il\theta},
\end{equation}
where $l$ is the orbital angular momentum quantum number. This leads to the Bessel equation
\begin{eqnarray}
R''(\rho)+\frac{1}{\rho}R'(\rho)+\left[\frac{\nu^2}{\rho^{2}}-k^{2}\right]R(\rho)=0, \label{bessel eqn}
\end{eqnarray}
with 
\begin{equation}
k^{2}=-\frac{2m\mathcal{E}}{\hbar^{2}} \label{k}
\end{equation}
and $\nu$ given by
\begin{equation}
\nu^{2}=\frac{(p^{2}-1)d^{2}}{48\pi}-(lp)^{2}.
\end{equation}
Notice that the condition which may lead to  bound state solutions depends on the balance between the electric dipole intensity $d$ and the angular momentum corrected by the parameter $p$. So, we are interested in those low angular momentum states where $\nu^2 > 0$. 

Now we do the following transformation in the radial equation (\ref{bessel eqn}):
\begin{equation}
S(\rho)=\rho^{\frac{1}{2}}R(\rho).
\end{equation}
The radial equation is transformed to 
\begin{equation}
S''(\rho)+\left[\frac{\nu^2+\frac{1}{4}}{\rho^{2}}-k^{2}\right]S(\rho)=0, \label{S-equation}
\end{equation}
Also, for convenience, we make $\rho=x/k$ and $S(\rho)=S(x/k)=u(x)$
where $x$ is a adimensional quantity. Eq. (\ref{S-equation}) becomes now
\begin{eqnarray}
u''(x)+\left[\frac{\nu^2 +\frac{1}{4}}{x^{2}}-1\right]u(x)=0, \label{xeqn}
\end{eqnarray}
where the derivatives are now taken with respect to $x$. 
The solution to equation (\ref{xeqn}) is given in terms of the modified Bessel function of third kind and imaginary order $K_{i\nu}(x)$  \cite{watson}
\begin{equation}
u(x)= \sqrt{x}K_{i\nu}(x),
\end{equation}
with the range $[ka,\infty ]$, where $a$ is the radius of the string. Therefore,  we require 
\begin{equation}
u(ka)= \sqrt{ka}K_{i\nu}(ka)=0.
\end{equation}

In order to find the zeros of the Bessel function $K_{i\nu}(x)$ we look at the series expansion \cite{gil}
\begin{equation}
K_{i\nu}(x)=\sum_{j=0}^{\infty} c_{j}f_{j},
\end{equation}
where
\begin{equation}
c_{j}=\left(\frac{x^2}{4}\right)^{j}\frac{1}{k!} \label{c}
\end{equation}
and
\begin{equation}
f_{j}=\frac{\pi}{2\sin (i\nu\pi )}\left[\frac{(x/2)^{-i\nu}}{\Gamma(j+1-i\nu)}-
\frac{(x/2)^{i\nu}}{\Gamma(j+1+i\nu)}\right]. \label{f}
\end{equation}
The expansion is justified because of the smallness of the string radius and also because $p$ is very close to $1$. 

Now, the above equations (\ref{c}) and (\ref{f}) yield 
\begin{eqnarray}
c_{0}&=&1,  \nonumber \\
f_{0}&=& \sqrt  {\frac{\pi}{\nu\sinh (\pi\nu)}}\sin \left[\nu \ln (x/2)+\nu\gamma \right],
\end{eqnarray}
where $\gamma$ is Euler's constant.
This leads  to the following expansion
\begin{equation}
K_{i\nu}(x)\approx  \sqrt  {\frac{\pi}{\nu\sinh (\pi\nu)}}\sin \left[\nu \ln (x/2)+\nu\gamma \right] + O(\nu).
\end{equation}
Therefore, for small $x$ and $\nu$, the zeros of the Bessel $K_{i\nu}(x)$ function are given by
\begin{equation}
x_{n}=2e^{-n\pi / \nu-\gamma}[1+O(\nu)],
\end{equation}
where, $n=1,2,...,\infty$.
From $ka=x_n$ and from equation (\ref{k}) we get the energy levels
\begin{equation}
\mathcal{E}=-\frac{2\hbar^2 e^{-2n\pi / \nu-2\gamma}}{m a^2}[1+O(\nu)].
\end{equation}
Notice that the spectrum has a point of accumulation at zero and that, if the radius $a$ of the string is taken to zero there, is no lower bound for the energy levels, which will range from minus infinity to zero.  Fortunately, a physical string must have a non-vanishing core, providing a natural regularization for this problem.

In summary, in this work we analyze a  non-relativistic problem in which an electric dipole interacts with a cosmic string via a $1/\rho^{2}$ effective potential. We have shown that this  potential under a specific condition leads to bound states. This potential is nevertheless pathological in the sense that the bound states have no lower bound, ranging from minus infinity to zero, making these bound states unstable. This difficulty  comes from the assumption of an infinitely thin cosmic string. Actually, a more realistic  model for the cosmic string presumes a finite string radius. This provides a natural cutoff for the $1/\rho^{2}$ singularity and gives a well defined energy spectrum for the problem. Finally, the study of the spectrum of  dipoles bound to a cosmic string may be a useful tool for the interpretation of astrophysical data in the search for cosmic strings.

{\bf Acknowledgments}\\
\noindent 
We would like to thank  CNPq, CAPES (PROCAD), FAPESB and PROINPE from Universidade
Estadual de Feira de Santana for financial supports.

\end{document}